\documentclass[useAMS,usenatbib]{mn2e}

\usepackage{epsfig}
\usepackage{amsmath}
\usepackage{amssymb}
\usepackage{ifthen}
\usepackage{txfonts}
\usepackage{rotating}
\usepackage{url}
\usepackage{varioref}
\usepackage{verbatim}
\usepackage{latexsym}
\usepackage{graphicx}
\usepackage{subfigure}

\newcommand\araa{\rmfamily{ARA\&A}}%
\newcommand\apj{\rmfamily{ApJ}}%
\newcommand\apjl{\rmfamily{ApJ}}%
\newcommand\aap{\rmfamily{A\&A}}%
\newcommand\mnras{\rmfamily{MNRAS}}%
\newcommand\prd{\rmfamily{Phys.~Rev.~D}}%
\newcommand\pasj{\rmfamily{PASJ}}%
\newcommand\nat{\rmfamily{Nature}}%
\newcommand\aplett{\rmfamily{Astrophys.~Lett.}}%
\title[3D Stability of Relativistic Jets from Black Holes]
{Stability of Relativistic Jets from Rotating, Accreting Black Holes via Fully Three-Dimensional Magnetohydrodynamic Simulations}

\author[J.~C. McKinney, \& R.~D. Blandford]
   {Jonathan C. McKinney,$^1$$^2$\thanks{\hbox{E-mail: jmckinne@stanford.edu~(JCM);} \hbox{rdb3@stanford.edu~(RDB);} }
    Roger D. Blandford$^1$\footnotemark[1] \\
  $^1$Department of Physics and Kavli Institute for Particle Astrophysics and Cosmology, Stanford University, Stanford, CA 94305-4060, USA \\
  $^2$Chandra Fellow }

\begin{document}
\date{Accepted 2009 January 14.  Received 2009 January 14; in original form 2008 December 4}
\pagerange{\pageref{firstpage}--\pageref{lastpage}} \pubyear{2009}
\maketitle

\label{firstpage}
\begin{abstract}

Rotating magnetized compact objects and their accretion discs can
generate strong toroidal magnetic fields driving highly magnetized plasmas
into relativistic jets.  Of significant concern,
however, has been that a strong toroidal field in the jet should be highly unstable
to the non-axisymmetric helical kink (screw) $m=1$ mode leading to rapid disruption.
In addition, a recent concern has been that the jet formation process itself
may be unstable due to the accretion of non-dipolar magnetic fields.
We describe large-scale fully three-dimensional
global general relativistic magnetohydrodynamic simulations of
rapidly rotating, accreting black holes producing jets.
We study both the stability of the jet as it propagates and the stability of the jet formation process
during accretion of dipolar and quadrupolar fields.
For our dipolar model, despite strong non-axisymmetric disc turbulence,
the jet reaches Lorentz factors of $\Gamma\sim 10$ with opening half-angle $\theta_j\sim 5^\circ$ at $10^3$ gravitational
radii without significant disruption or dissipation with only mild substructure dominated by the $m=1$ mode.
On the contrary, our quadrupolar model does not produce a steady relativistic ($\Gamma\gtrsim 3$) jet
due to mass-loading of the polar regions caused by unstable polar fields.
Thus, if produced, relativistic jets are roughly stable structures
and may reach up to an external shock with strong magnetic fields.
We discuss the astrophysical implications of the accreted magnetic geometry
playing such a significant role in relativistic jet formation, and we outline avenues
for future work.

\end{abstract}


\section{Introduction}

Astrophysical jets were discovered by Heber Curtis in 1917, who described M87's jet as
``a curious straight ray ... connected with the nucleus'' \citep{curtis1918}.
The M87 jet is the most well-studied of all jets
associated with active galactic nuclei (AGN) (e.g. \citealt{bm54,bk79a,rm93}).
M87's jet structure has been observed down to tens of gravitational radii
of the putative black hole (BH) (e.g. \citealt{junor1999,kovalev2007,ly2007})
with impressive animations created \citep{walker2008}.
Jets have now also been observed in many other AGN/blazars \citep{bp84},
in neutron star and BH x-ray binaries \citep{mr99,fender04},
in Herbig-Haro objects (e.g. \citealt{konigl82}),
and are required for gamma-ray bursts (GRBs) (e.g. \citealt{piran2005}).
Challenges include explaining
the jet formation process,
the stability of jet formation and jet propagation,
how jets accelerate and collimate,
and how jets obtain their composition and substructure both near and far from the central object.
Jet studies are complicated by the system's environment,
such as how a jet must drill through a massive envelope in the collapsar model, while
FRII jets extend up to hot spots.
For quasar systems, jets play an important role
in limiting BH mass growth (e.g. \citealt{dsh05})
and driving hot bubbles that limit cooling flows (e.g. \citealt{mcn05}).  However,
the efficiency of the energy-momentum transfer remains unknown and
probably depends on jet structure and stability.

The most universally applicable jet paradigm involves some form
of magnetic-driving with strong toroidal fields that form, accelerate,
and (internally) collimate jets via magnetized accretion discs
\citep{lb69,br74,lovelace76,bp82} or accreting, rotating BHs (\citealt{bz77}, BZ).
This paradigm was bolstered by the realization that accretion can be driven
by magnetorotational turbulence generating a strong magnetic field
\citep{bh98}.
Especially if discs are thick near BHs,
then stronger jets and winds are driven by either
stronger turbulent magnetic fields for the otherwise same mass accretion rate
(e.g. \citealt{meier01,miller06}) or by
large-scale fields advected from large radii (see \S3 \& \S4 in \citealt{lop99}).
A magnetic field may preserve jet composition against entrainment \citep{rosen99}.
The variations in magnetic field strength and BH spin
may explain the diversity of jet systems like FRI/FRII's \citep{meier99},
although rotation measures imply unexpected field orientations \citep{zt05} and
simple models of decelerating jets fit FRIs (e.g. \citealt{laing06}).
Soltan efficiency (and other) arguments suggest quasars contain BHs that are
rapidly spinning (e.g. \citealt{gsm04}) and maybe
maximally spinning (e.g. \citealt{allen06}).
The intrinsic interest (and cosmological importance) of jets
motivates testing whether the magnetic paradigm can explain
their observed structure and stability.

Now roughly 90 years after Heber Curtis's discovery, the {\it straightness}
of many observed jets remains as their most inexplicable feature
given, e.g., fusion devices show strong toroidal fields
are violently unstable to helical kink (screw) modes (e.g. \citealt{bateman78}).
Astrophysical jet stability research has revealed
a large number of modes (e.g. \citealt{kad66})
that can be unstable including ``reflection'' resonant modes \citep{pc85},
Kelvin-Helmholtz (KH) modes (e.g. \citealt{ferrari78} and references therein),
and current-driven modes \citep{benford81}.
With perturbations of the form $e^{i(m\phi + n z + l R - \omega t)}$,
a nearly universal result from these simplified models is that the $m=1$ kink
mode is the most dangerous mode that could result in complete disruption and dissipation.

Even if simplified jet models are kink mode unstable,
they may be stabilized by introducing gradual shear (e.g. \citealt{mhn07} and references therein),
an external wind \citep{hh03}, sideways expansion \citep{rh00},
and relativistic bulk motion.
For some AGN jets, observations support a lack of significant
dissipation during propagation \citep{sambruna06}.
If unstable, however, jets can be a source of heating, radiation, and high-energy particles
due to shocks (e.g. \citealt{bk79a}),
reconnection (e.g. \citealt{ds02,lyutikov06,gs06}),
viscous shear,
turbulent cascade (e.g. \citealt{begelman98}),
and a break-down of the ideal single-component fluid approximation \citep{trussoni1988}.

For magnetized jets, the current-driven screw ($n>0, m=1$) mode is potentially
most disruptive.  For cylindrical force-free equilibria one obtains
the Kruskal-Shafranov (KS) instability criterion
\begin{equation}\label{kscrit}
-\frac{B^\phi}{B^p}>\frac{2\pi R}{r} ,
\end{equation}
where $B^\phi$ and $B^p$ are the toroidal and poloidal field strengths,
$R=r\sin\theta$ is cylindrical radius, and $r$ is poloidal extent.
This suggests jets are unstable beyond the Alfv\'en surface
where $B^\phi\gtrsim B^p$ and $r\gtrsim R$, located at
only $r\lesssim 10M$ (in this Letter, $G\equiv c\equiv 1$)
for rotating BHs or accretion discs.
The KS criterion implies jets go unstable before accelerating to relativistic
($\Gamma\gtrsim 3$) speeds as likely only after $r\sim 100M$ \citep{M06a},
and the KS criterion probably cannot explain some FRIIs extending to $r\sim 10^7M$.

Advanced linear stability analyses from normal mode and extremal energy
arguments for simplified cylindrical jets are often based upon restrictive assumptions,
which has lead researchers to suggest that jets are
violently unstable \citep{begelman98,lyub99,li00},
mildly unstable (e.g. \citealt{lery00}),
or even stable (e.g. \citealt{ip96}) to the screw mode.
\citet{tmt01} found that relativistic field rotation of freely
expanding solutions \citep{M06a,M06c,narayan2007} allows jets to be unstable
only if {\it both} the KS criterion and their criterion,
\begin{equation}\label{tmtcrit}
-\frac{B^\phi}{B^p}>\frac{R\Omega_F}{c} ,
\end{equation}
are satisfied, where $\Omega_F$ is the field line rotation frequency
and $c$ is the speed of light.  This implies jets are marginally
stable until a strong external medium interaction.
Their analysis is suggestive, but it only strictly applies inside,
not through, the Alfv\'en surface.  So far, no sufficiently general analytical screw stability
analysis has been performed for magnetically-dominated relativistic jets.

Analytical approaches become intractable for more realistic jets.
It remains difficult to compare theory with observations (e.g. \citealt{worrall2007})
and laboratory experiments (e.g. \citealt{ciardi08}).
Primarily, numerical magnetohydrodynamical (MHD) simulations
have proven useful to study realistic jet models.  Simulations
range from injecting an arbitrary jet from a surface inlet
(e.g. \citealt{nm04,zhang04,leismann05,kvkb08,mso08})
to injecting a jet from an unresolved Keplerian disc (e.g., \citealt{tmn08}),
and to evolving both the disc and jet (e.g.,
\citealt{hbs01,mg02,mg04,ks05,M05,hk06,M06a,mn07a,mn07b,km07}).
Advanced 3D MHD simulations that inject jets from an inlet
find that KH kink modes are stabilized by sheaths around the jets \citep{mhn07}
and that even non-relativistic screw modes saturate before causing magnetic dissipation \citep{mso08}.
More realistic simulations are crucial since
analytical experience suggests free parameters in jet-injection
simulations probably play a significant role.
In particular, only global simulations allow a stability
study of the actual jet formation process in the presence of
disc turbulence and different global field geometries.
Accretion of small quadrupolar field loops was already shown to
degrade the jet \citep{mg04,beckwith08}, but this could be due to their choice of
starting with small field loops in the disc with numerical dissipation not allowing the development
of a large-scale quadrupolar field.

\section{Numerical Model}

We perform fully 3D global general relativistic MHD (GRMHD)
simulations starting with an equilibrium matter torus,
whose angular momentum is aligned with the BH (Kerr metric) spin.
To facilitate comparisons, we follow \citet{M06a} and choose a
torus pressure maximum at $r=12M$, inner edge at $r=6M$, and adiabatic index $\gamma=4/3$
giving disc thickness $\delta\theta\sim \pm 0.3$.  For BH spins of $a/M\gtrsim 0.4$,
simulations of such tori are qualitatively similar \citep{mg04}.
We choose all models to have $a/M=0.92$
(hole angular frequency, $\Omega_H=a/(2Mr_+)\approx 0.33 M^{-1}$, with horizon radius, $r_+$)
such that the BH is in spin equilibrium for our disc thickness \citep{gsm04}.
We use the conservative unsplit 3D GRMHD code HARM \citep{gmt03}, Kerr-Schild
coordinates, $4$th-order interpolation and $4$th-order Runge-Kutta \citep{M06a},
a robust inversion scheme \citep{mm07}, a staggered field scheme (McKinney et al., in prep.),
and other advances \citep{M06b,tmn07}.

We consider both dipolar and quadrupolar field geometries.
The dipolar model starts with a single field loop within the torus as in \citet{M06a}.
Dipolar models correspond to the most-often simulated jet (or jet+disc) model
were the current sheet is assumed to be at (or develops near) the equator.
To give the quadrupole geometry the best chance of producing a jet,
we study a {\it large-scale} quadrupolar field with vector potential $\phi$ component
\begin{equation}
A_{\rm quadrupole} = A_{\rm dipole} \cos\theta ,
\end{equation}
using a paraboloidal-like potential
given by
\begin{equation}
A_{\rm dipole} = (1/2) [(r+r_0)^\nu f_- + 2Mf_+(1-\ln(f_+))] ,
\end{equation}
where $f_- = 1-\cos^\mu\theta$, $f_+=1+\cos^\mu\theta$, $\nu=3/4$,
$\mu=4$, $r_0=4$, and applies for $\theta<\pi/2$
and for $\theta>\pi/2$ when letting $\theta\rightarrow\pi-\theta$.
In this model, current sheets form above and below the equator.
From prior GRMHD simulations, we expect primarily the initial field's
multipole order to be important, and particular model parameter values
should be unimportant once a quasi-steady state is reached.
All models have initial gas pressure per unit magnetic pressure of $\approx 100$ at the equator in the disc.
We allow the comoving magnetic energy per rest-mass energy up to only $100$
during mass evacuation near the BH (see floor model in \citealt{M06a}).

Spherical polar, not Cartesian, coordinates are used since preferred for rotating jets.
Our fiducial models have resolution
$256\times 128\times 32$ in $r\times\theta\times\phi$,
with non-uniform grid as in \citet{M06a}, except $R_0=0$ and $n_r=1$ in their equation (18).
Based upon code tests, our $2$nd-order monotonized central limiter
scheme would require roughly $4\times$ the per-dimension resolution to
obtain the accuracy our $4$th-order scheme by the end of the simulation.
Unlike prior 3D GRMHD simulations, the grid warps to resolve the disc at small radii
and follows the collimating jet at large radii giving roughly $3\times$ more angular
resolution at large radii.  Hence, compared to any scheme similar to the original $2$nd-order HARM scheme,
our effective resolution is roughly $1024\times 1536\times 128$.
Unlike most 3D GRMHD simulations (e.g. \citealt{beckwith08}),
we include the full $\Delta\phi=2\pi$ extent as required to resolve
the $m=1$ mode and include the full $\Delta\theta=\pi$ extent (no cut-out at poles).
As \citet{fragile07}, we use transmissive (not reflecting) polar boundary conditions.  As they state,
the singularity need not be treated specially for centered quantities in a finite-volume scheme.
Our field is staggered, and the polar value of $B^\theta$ is evolved by using the analytical limit of the
finite volume induction equation at the pole such that angular-dependent area factors cancel (McKinney et al., in prep.).
Coordinate directions twist at the pole leading to some dissipation,
but this is significantly reduced by our $4$th-order scheme that well-resolves up to $m=4$ with $32$ $\phi$ cells.
At the inner torus edge, cells have aspect ratio 1:5:10 and the fastest-growing
magnetorotational mode is resolved with $6$ cells, as sufficient \citep{shafee08}.
We also studied resolutions of $128\times128\times32$,
$128\times64\times32$, and $128\times64\times16$; the
jet's Fourier $m=1,2,3$ power is converged to $20\%$.
Using $128$ angular cells and staggered field scheme were required for MHD
jet invariants to be conserved to $\lesssim 10$\%, which is evidence
of an accurate solution \citep{tmn08}.

Most disc+jet simulations do not evolve
to large enough radii to resolve a highly relativistic jet.
For magnetically-dominated paraboloidal jets, the maximum Lorentz factor at
large radii is
\begin{equation}
\Gamma\approx 0.3\left(\frac{r}{M}\right)^{0.5} ,
\end{equation}
\citep{tmn08}.
We choose an outer box radius of $10^3M$ as required to reach $\Gamma\sim 10$.
All simulations ran a duration of $5000M$, which is $192$
orbits at the inner-most stable circular orbit (ISCO)
($r_{\rm ISCO}\approx 2.2M$) and $50$ orbits at the initial
inner torus edge.  The accretion rate of mass ($\dot{M}$), energy,
and angular momentum are roughly constant with radius
out to $r\sim 10M$ by $t\sim 3000M$,
indicating the disc has reached a quasi-steady state.
The slow/contact modes for the jet move with $v/c\gtrsim 0.2$,
so the jet is beyond the box by $t=5000M$.  We report
many results at $t\sim 4000M$ since this is before the jet
partially reflects off the outer box.

\section{Results}

\begin{figure}

\includegraphics[width=3.3in,clip]{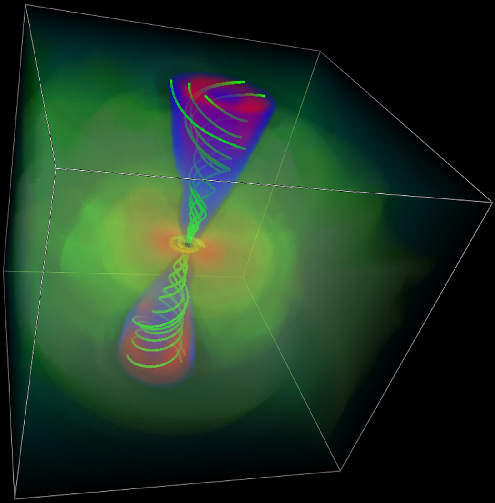}

\caption{For dipolar model, shows inner $\pm 100M$ cubical region
with BH, accretion disc (pressure, yellow isosurface),
outer disc and wind (log rest-mass density, low green, high orange, volume rendering),
relativistic jet (Lorentz factor of $\Gamma\lesssim 4$, low blue, high red, volume rendering),
and magnetic field lines (green) threading BH.
Despite non-axisymmetric turbulence, polar magnetically-dominated jets
are launched by the BZ effect.}
\label{disc} \end{figure}

\begin{figure}

\includegraphics[width=3.3in,clip]{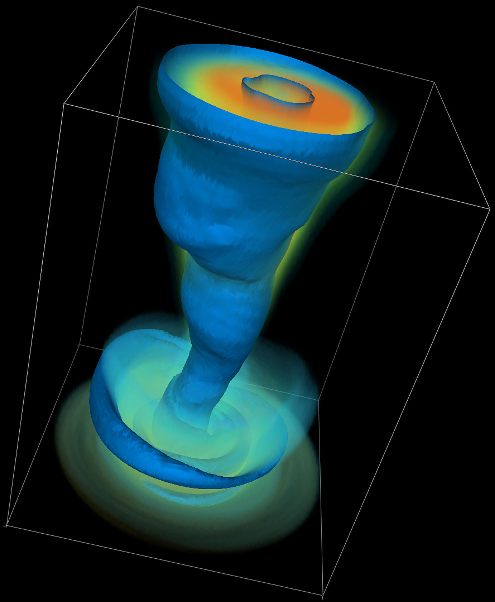}

\caption{For dipolar model, shows accreting BH
generating relativistic jet (only one side shown)
vertically out to $10^3M$ within $\theta\approx\pm 20^\circ$
($350M\times 350 M$) at $t=4000M$.
Shown are outskirts of disc and wind (log internal energy density, cyan volume rendering),
outer/inner boundary of perturbed jet and fragments of disc wind ($R B^\phi$, blue isosurface),
and relativistic jet (Lorentz factor of $\Gamma\lesssim 10$, orange volume rendering)
collimated within half-angle $\theta_j\approx 5^\circ$.
Despite perturbations, the jet is a stable structure. }
\label{jet} \end{figure}

The fiducial dipole model is overall similar to prior
2D simulations \citep{mg04,M06a}.  The BH-driven polar jet
survives in a non-dissipated state to large radii.
Each polar, magnetically-dominated jet at
$r_+,10,10^2,10^3M$ has constant electromagnetic luminosity of
$L_j\approx 0.01\dot{M}c^2$, with only a small secular drop as $\Gamma$ increases.
This value is similar to higher resolution 2D simulations \citep{mg04}.
The total (disc+jet+wind) electromagnetic output peaks at $r\approx 10M$,
but disc power is dissipated so does not survive at large radii \citep{mn07a}.
Figure~(\ref{disc}) shows the inner $\pm 100M$ cubical region and
Figure~(\ref{jet}) shows out to $z=10^3M$ by $t=4000M$.
The figures show the disc wind and relativistic jet generated by the rotating
BH and magnetized, turbulent accretion disc.  The jet is
roughly stable out to $z=10^3M$ reaching $\Gamma\sim 5-10$.
Figure~(\ref{jet}) shows the kinked polar jet structure of the
poloidal current, $R B^\phi$, capable of driving screw instabilities.
We measure the Fourier power within the jet region defined by
magnetic energy per rest-mass energy, averaged for all $\phi$, greater than one.
At large distances the $m=1,2,3,4$ powers relative to $m=0$
are $7\%,1\%,0.7\%,0.6\%$ in magnetic energy, $6\%,4\%,0.5\%,0.2\%$ in Lorentz factor,
roughly $37\%,7\%,3\%,4\%$ in both rest-mass density ($\rho_0$) and $R B^\phi$,
and $20\%,13\%,7\%,6\%$ in internal energy density. Both $\rho_0$ and $R B^\phi$ reach
$m=1$ power of $100\%$ in the jet next to the outer disc edge at $r=20M$.
There is no indication of growth beyond perturbations induced by the disc turbulence,
which appears to be the primary origin of jet substructure.

Now we discuss our fiducial large-scale quadrupole model.
GRMHD simulations show that no strong jet emerges
due to the accretion of higher multipole moments put initially within the disc \citep{mg04,mn07a,mn07b,beckwith08}.
In our fully 3D simulations, even putting in an initial
large-scale quadrupolar field leads to no strong jet
once equatorial symmetry is broken by $t\sim 2500M$.
While the polar field strength is similar to that in the dipole model when the field threads the horizon,
the polar regions are mass-loaded when tearing coronal current sheets
eject polar field to slightly larger radii.
Then, magnetic pressure no longer balances against the
low angular momentum disc material that moves into the polar region.
The coronal-polar field is accreted and ejected throughout the simulation,
which leaves no time allowed for the funnel to drain.
This leads to order unity magnetic energy and internal energy per particle rest-mass energy,
which is insufficient to generate a highly relativistic ($\Gamma\gtrsim 3$) jet.
Similar results by \citet{beckwith08} using a non-energy-conserving code suggest
detailed thermodynamics do not control this process.
At late times, (total and polar) BH electromagnetic power are negative.
The polar regions have both inflows and outflows,
and there is only a disc-mass-loaded wind with an electromagnetic power output
per pole at large radii of $L_w\approx 0.002\dot{M}c^2$, which is significantly
less powerful than the dipole model.
Similar as the disc wind in the dipolar model, the outflow
has a weak disorganized poloidal field and a more organized toroidal field
stronger by factors of typically $10-40$ both near the BH and at large radii.
Corresponding 2D simulations show less drastic, but comparable, degradation of the jet.
A thinner disc may not allow as much mass-transfer to the poles, but
thinner discs have weaker turbulent fields and inward advection of strong ordered
field may not be possible for thin discs.
Also, higher resolutions may lead to less vigorous reconnection
or may show a more narrow, polar jet still emerges.

\section{Discussion}

We have performed fully 3D global GRMHD simulations of accreting, rapidly rotating
BHs and found that dipolar fields near BHs can launch
magnetically-dominated, relativistic ($\Gamma\gtrsim 3$) jets that
survive to $10^3M$ without significant disruption or measurable dissipation.
Disc turbulence appears to be the primary cause of
jet substructure that is dominated by the $m=1$ mode,
which has no measurable growth within the jet.
Prior work applying a form of the Kruskal-Shafranov criterion
(solution for non-relativistic, cylindrical equilibria)
to highly magnetized relativistic flows (e.g. \citealt{lyutikov06,gs06}),
needs to be reevaluated to consider the stabilizing effects of
field rotation, gradual shear, a surrounding sheath, and sideways expansion as
present in the simulations.
Unlike dipolar fields, quadrupolar fields near BHs
lead to only weak, turbulent outflows and negligible magnetically-dominated polar
regions and no relativistic ($\Gamma\gtrsim 3$) jets.
Since our simulations with relativistic jets have no current sheets within the jet,
reconnection may not be an important source of
dissipation unlike assumed by some models (e.g. \citealt{ds02}.)

These and prior GRMHD simulation results suggest that a rotating ($a/M\gtrsim 0.4$) BH
is a necessary, but not sufficient, condition to produce a
highly relativistic ($\Gamma\gtrsim 3$) jet.  In addition, one requires
the accreted magnetic field to be mostly dipolar, rather than higher-order,
so a dipolar field threads the region near the BH (see also \citealt{nia03}).
This might explain various observations, such as the dichotomy of
FRI and FRII systems.  FRI's are found in rich clusters, are two-sided so weakly relativistic,
and have dissipative emission near the core.  FRII's are found in poor groups
or isolated, are one-sided so more relativistic, are more powerful, and dissipate
little till the radio lobe \citep{ol94}.
The FRI/FRII dichotomy may then be due to the complexity of the environment (e.g. through hierarchical merging)
controlling the field multipole structure.
Then, FRII systems are primarily BH-driven able to pierce through an ambient medium,
while FRI systems are those mostly driven by the
broader, dissipative, magnetically-disordered disc wind with $\Gamma\lesssim 3$
that one expects to be more easily entrained, slowed, and disrupted,
as consistent with observations \citep{laing06}.
Radial structure (e.g. arcs and knots) could be due to
accretion switching between dipolar and higher-order multipoles.
For M87, there could be a dark or boosted relativistic spine
with the slower, dissipative disc wind producing emission
on scales within several parsecs \citep{kovalev2007}.
For SrgA*, no jet may emerge because of accretion
from various stellar clusters generating a dominant non-dipolar field
\citep{ncs07}.  For x-ray binaries,
jets in the low-hard states could be driven by dipolar fields
that could even accumulate to the point of lowering accretion rates \citep{ina03},
intermediate to soft states could involve higher-order multipole moments,
and transient jets from the hard-to-soft transitions could occur
due to dissipation of the dipolar component.
For GRBs, the BH-disc system may be required to be highly symmetric to maintain
a strong dipolar field to produce an ultrarelativistic jet.
That ordered poloidal field must be accreted
assumes no dynamo exists for generating a baryon-pure,
large-scale poloidal field from disorganized field \citep{beckwith08}.

Future jet studies should consider the effects of much higher resolutions,
misaligned BH-disc accretion (since misaligned systems may more readily produce non-dipolar fields),
larger radii of $10^7M$ for AGN and $10^{12}M$ for GRBs
(to determine very large-scale stability and to obtain larger $\Gamma$),
resistivity and viscosity,
disc radial extent (that limits the terminal Lorentz factor since the
lack of the disc and supportive disc wind allows the jet
to become monopolar and so accelerate inefficiently),
disc thickness (that can control the strength of turbulent or advected field),
other magnetic field geometries (including with non-zero net helicity),
BH spin (especially very low and very high),
cooling (such as neutrino cooling in collapsar discs),
and the presence of an extended massive envelope as in the collapsar model
(freely expanding outflows simulated here
apply to a late phase after the jet drills through the envelope).
Future studies should also do a quantitative analysis of the modes
within the jet to identify which mode types are present.
The simulated jet can be used as a
well-motivated background state for future linear
perturbation analyses, parameter searches,
and synchrotron and inverse Compton maps for, e.g., VLBI, Chandra, and Fermi.

\section*{Acknowledgments}

We thank Ramesh Narayan and Alexander Tchekhovskoy for stimulating discussions.
Simulations were run on the TACC Lonestar and KIPAC Orange clusters.
Support was provided by NASA's Chandra Fellowship PF7-80048 (JCM),
NSF grant AST05-07732 (RDB), SciDAC grant DE-FC02-06ER41438 (JCM \& RDB),
and DOE contract DE-AC02-76SF00515 (JCM \& RDB).

\label{lastpage}

\end{document}